\documentclass[vecphys]{svmult}

\usepackage{makeidx}         
\usepackage{graphicx}        
\usepackage{multicol}        
\usepackage[bottom]{footmisc}

\makeindex             

\begin{document}

\title*{Jets in Active Galactic Nuclei}

\author{Alan P. Marscher}

\institute{Institute for Astrophysical Research, Boston University,
725 Commonwealth Ave., Boston, MA 02215, USA
\texttt{marscher@bu.edu}}

\maketitle

\abstract{
The jets of active galactic nuclei can carry a large
fraction of the accreted power of the black-hole system into
interstellar and even extragalactic space. They radiate profusely
from radio to X-ray and $\gamma$-ray frequencies. In the most extreme
cases, the outward flow speeds correspond to high Lorentz factors
that can reach 40 or more. This chapter displays images at
various wavebands as well as light curves and continuum spectra that
illustrate the variability with location, time, and frequency of the
emission from compact, parsec- and subparsec-scale jets. It presents a
physical framework for investigating many aspects of the structure and
dynamical processes from such data.
}

\section{Introduction}
\label{sec:1}

Since accretion onto stellar-mass black holes is often accompanied by a pair of jets
emanating from the rotational poles, we have the right to expect an even more
spectacular brand of jets in active galactic nuclei (AGN). Indeed, AGN jets were the first
to be observed. These systems,
with their supermassive black holes of millions to billions of solar masses, can produce
ultraluminous jets that bore their way through the interstellar medium and into
intergalactic space, ending in huge, billowy lobes punctuated by hotspots. Even the
less luminous versions disturb the gas in their host galaxies. During the epoch of galaxy
formation, this may have controlled the infall of gas so that the central black hole and
galactic bulge grew in step with each other.

The wide variety of AGN is reflected in the diversity of their jets. These range from
relatively slow, weak, and poorly collimated flows in Seyfert galaxies to strong jets with
relativistic speeds in Fanaroff-Riley\index{Fanaroff-Riley} (FR)~I radio galaxies and BL Lacertae
(BL~Lac) objects to the most luminous, highly focused, and relativistic beams in FR~II
radio galaxies and radio-loud quasars. The
reason for this dichotomy is unclear, although it probably relates to the galactic
environment, with Seyfert galaxies usually of spiral morphology, FR~I sources hosted by
giant elliptical galaxies in rich clusters, and FR~II objects in elliptical galaxies
lying in somewhat less dense groupings. This might influence the rate of accretion of gas
or spin of the black hole, one (or both) of which could be the factor that determines
how fast and well-focused an outflow is propelled in the two polar directions.

The most extreme jets with highly relativistic flow velocities are in fact the best-studied
variety. This is because relativistic beaming of the radiation amplifies the brightness
of these jets so that they can be prominent even at relatively low luminosities
if one of the jets points within several degrees of the line of sight. Emission from
the jet often dominates the spectral energy distribution of the source in this case,
so that observations across the electromagnetic spectrum serve to define the properties
of the jet. For this reason, and because one often learns the most by studying
extreme cosmic phenomena, this chapter will emphasize relativistic jets pointing
nearly at us. AGN with such jets are termed ``blazars.''

Because of space limitations, this chapter discusses only a fraction of the
phenomenology and physics of AGN jets. The interested reader should consult other
sources of information, especially \cite{BBR84,Hughes91,Kro99,DY02}.
\begin{figure}
\centering
\includegraphics[height=6cm]{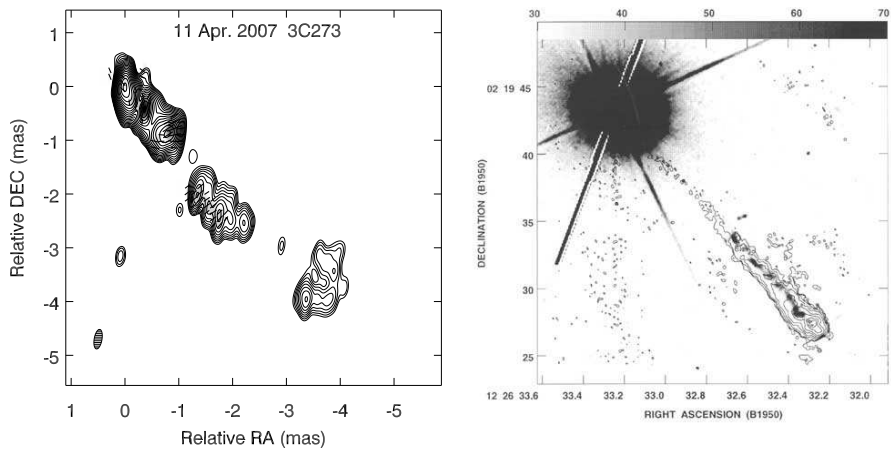}
\caption{Example of an AGN jet, in the quasar 3C~273\index{3C~273}. {\bf (Left)} on parsec scales
at radio frequency 43 GHz; {\bf (right)}  on kiloparsec scales at radio frequency
1.7 GHz (contours) and optical wavelengths centered on 606 nm (greyscale). The core
(brightest feature, at the northeast end) of the parsec-scale jet is located very
close to the center of the overexposed nucleus in the upper-left corner of the
right-hand panel. The scale of the image on the right is 4000 times larger than
that on the left. Contours of the radio images in this and subsequent figures are in
increments of a factor of 2. At the redshift of 0.158, $1'' = 2.7$
kpc. {\bf Left:} author's image, derived from data obtained with the Very Long Baseline
Array (VLBA); {\bf right:} from \cite{Bah95}, from data obtained with the NASA's Chandra
X-ray Observatory and the Very Large Array (VLA). The VLBA and VLA are instruments of
the National Radio Astronomy Observatory, a facility of the National Science Foundation
operated under cooperative agreement by Associated Universities, Inc.}
\label{fig:1}
\end{figure}

\section{Observed Properties of Jets}
\label{sec:2}
The jets of blazars are indeed extreme, with kinetic powers that can exceed the
luminosities of entire galaxies and with flow speeds sometimes greater than
$0.999c$. This section contains a
brief overview of the characteristics of jets compiled from a wealth of imaging,
spectral, and multiwaveband monitoring and polarization data. Subsequent sections
will examine these properties in more detail from a physical standpoint.

\subsection{Images at Different Wavebands}
\label{sec:2.1}
Figure \ref{fig:1} shows an example of a jet on both parsec and kiloparsec scales,
while Figure 2 compares the X-ray and radio emission from a jet on kiloparsec scales. The
jet, which broadens with distance from the nucleus, is defined by knots of emission.
There is an overall likeness on the different scales, which indicates that jet flows are
roughly self-similar over several orders of magnitude in distance from the black hole.
This is easiest to understand if the jets are mainly fluid 
phenomena so that they are governed by the laws of gas dynamics and magnetohydrodynamics.

Images of jets contain a bright, nearly unresolved ``core'' at the upstream end.
This roughly stationary feature can represent a physical structure,
such as a standing shock wave, or simply the most compact bright emission that is
visible at the frequency of observation. In the latter case, the core can be at a
different position at different wavelengths. The remainder of the jet tends to be
knotty, although regions of smoothly varying intensity are sometimes present.
On parsec scales imaged by very long baseline interferometry (VLBI), we find that
many of the knots move, while other emission features appear stationary. Motions
probably occur on kiloparsec scales as well, but monitoring over very long times
is needed to detect such motions.

\begin{figure}
\centering
\includegraphics[height=7cm]{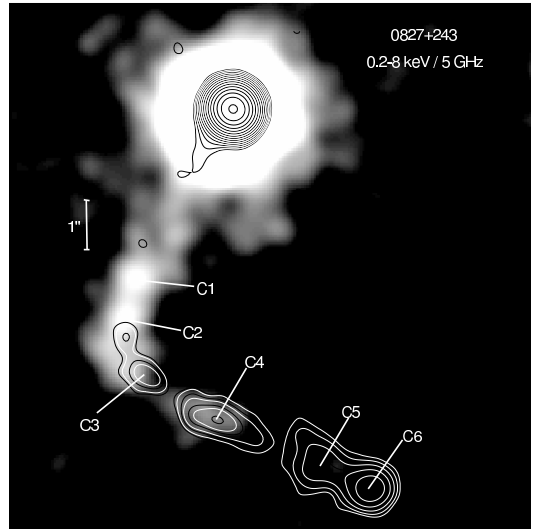}
\caption{Image of the jet of the quasar 0827+243 (OJ~248) at radio frequency
5 GHz (contours) and at X-ray photon energies (greyscale). The nucleus is the bright region
near the top. At the redshift (0.939) of this quasar, $1'' = 7.2$ kpc on the sky,
but $\sim 100$ kpc when de-projected. From \cite{JM04}}
\label{fig:2}
\end{figure}

\subsection{Motions inside Jets and Relativistic Beaming of Radiation}
\label{sec:2.2}
\begin{figure}
\centering
\includegraphics[height=14cm]{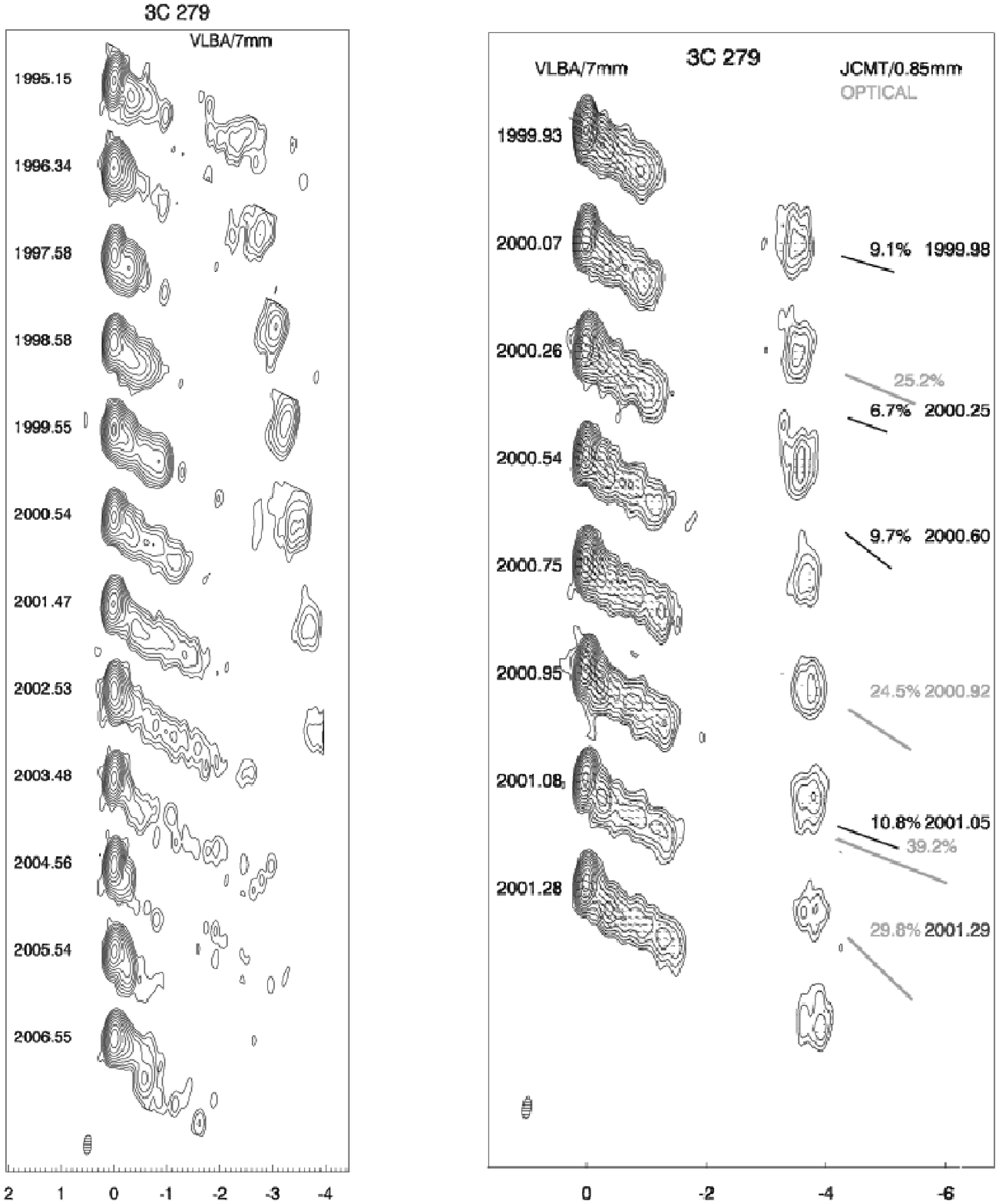}
\caption{{\bf (Left)} Twelve-year sequence of annual VLBI images of the quasar 3C~279\index{3C~279} at 43 GHz.
At the redshift of the quasar, 1 mas = 6.3 pc.
The feature at the narrow end of the jet, closest to the supermassive black hole, is referred to
as the ``core'' and is assumed to be stationary. The motion of the outlying knot has an
apparent superluminal speed of $\sim 10c$. Note the change in direction of the innermost
jet with time. {\bf (Right)} Sequence of roughly bimonthly
images of 3C~279 showing the systematic growth of the length of the inner jet
at a speed of $13c$. The linear polarization\index{polarization} at 43 GHz
is indicated by the sticks inside the contours, while the 230-350 GHz and optical
polarization are given on the right-hand side of the figure. Images from author and
S. Jorstad}
\label{fig:3}
\end{figure}

Strong evidence that some jets have relativistic flow velocities comes from
sequences of VLBI images that show apparent superluminal motions of bright knots
away from the core, as illustrated in Fig. \ref{fig:3}. The apparent motion
is faster than the actual speed $\beta c$ because the emitting plasma approaches us at
nearly the speed of light $c$. Therefore, it moves almost as fast as the radio waves
it emits, so that the arrival times at Earth are compressed relative to those in the
rest frame of the plasma. The apparent velocity in the plane of the sky is given by
\begin{equation}
v_{\rm app} = {{\beta c \sin\;\theta}\over{1-\beta \cos\;\theta}},
\end{equation}
where $\theta$ is the angle between the velocity vector and the line of sight.
The apparent velocity is a maximum at $\theta = \sin^{-1}(1/\Gamma)$, where
$\Gamma = (1-\beta^2)^{-1/2}$ is the Lorentz factor of the centroid of the
``blob'' of emitting plasma.

While the time scales involved make superluminal\index{superluminal} motion easiest to detect on parsec
scales, a superluminal feature has been seen at least 120 parsecs from the nucleus
in the relatively nearby radio galaxy M~87 \cite{CHS07}.

The relativistic motion produces a strong Doppler effect that increases the observed
frequency and decreases the time scales of variability by a factor
\begin{equation}
\delta = [\Gamma(1-\beta \cos\;\theta)]^{-1},
\end{equation}
which is termed the ``Doppler factor\index{Doppler factor}.'' It is a maximum of $(1+\beta)\Gamma$
when $\theta = 0$ and equals $\Gamma$ when $\theta = \sin^{-1}(1/\Gamma)$.
Because of relativistic aberration,
most of the emission is beamed into a cone with an opening half-angle
$\sim \Gamma^{-1}$, so that the
flux density $F_\nu$ of the plasma is boosted by a factor of $\delta^2$. Since the
frequency $\nu$ is also increased, if $F_\nu \propto \nu^{-\alpha}$, the flux density
is a factor of $\delta^{2+\alpha}$ higher than it would be if the velocity were
much less than the speed of light. This applies if the emission comes from a feature
in a steady state---a standing shock wave, for example. However, if the emitting plasma
moves, as does a superluminal knot, there is another factor of $\delta$ from the
time compression ($\Delta T \propto \nu^{-1} \propto \delta^{-1}$). We then have
\begin{eqnarray}
F_\nu \propto \delta^{2+\alpha}\nu^{-\alpha}&{\rm (steady~state)}\\
F_\nu \propto \delta^{3+\alpha}\nu^{-\alpha}&{\rm (moving~plasma)}.
\end{eqnarray}

Jets have basically conical shapes, with opening half-angles $\phi$ ranging from
$\sim 5^\circ$ to considerably less than $1^\circ$. A study of jets in 15 AGN indicates that
$\phi \propto \Gamma^{-1}$ \cite{J05}. This implies that acceleration of jets to the highest
velocities is related to their collimation toward a nearly cylindrical geometry.

\subsection{Polarization}
\label{sec:2.3}
Jets are usually linearly polarized at a level that can be as high as 50\% or greater for
the more extended features, but more typically a few percent in the core. On the other
hand, the circular polarization\index{polarization}, with few exceptions,
is generally less than 1\%. This is the signature of
incoherent synchrotron radiation, with relativistic electrons over a wide range of
pitch angles spiraling in a magnetic field. The field can be nearly uniform over the
resolution beam of the interferometer in the case of linear polarization of tens of percent.
Lower levels of polarization indicate a chaotic magnetic field, which can be described
roughly in terms of $N$ cells of equal size, each with uniform but randomly oriented
field. The degree of polarization is then
\begin{equation}
p = p_{\rm max}N^{-1/2} \pm p_{\rm max}(2N)^{-1/2},
\end{equation}
where $p_{\rm max} = (\alpha+1)/(\alpha+5/3)$ is usually in the range of 0.7-0.75
\cite{Burn66}.

The electric vector position angle (EVPA, or $\chi$) of linear polarization is
perpendicular to the
projected direction of the magnetic field if the emission is from optically thin
synchrotron radiation. However, relativistic motion aberrates the angles, an
effect that causes the observed value of $\chi$ to align more closely with the jet
direction than is the case in the plasma rest frame. Because of this, polarization
electric vectors observed to be roughly perpendicular to the jet axis correspond
to magnetic fields that lie essentially parallel to the flow direction, but EVPAs that
are modestly misaligned with the jet actually imply that the rest-frame magnetic field
is quite oblique to the jet axis.

Faraday rotation alters the EVPA as well. If the synchrotron emission from the jet
passes through a Faraday ``screen''---e.g., a sheath of thermal plasma surrounding the
jet---then the EVPA rotates by an amount
\begin{equation}
\Delta\chi = 7.27\times 10^4 [\nu_{\rm GHz}/(1+z)]^{-2} \int n_{\rm e} B_\parallel
ds_{\rm pc}~~{\rm radians},
\end{equation}
where $z$ is the redshift of the AGN, $n_{\rm e}$ is the electron density of the
screen in cm$^{-3}$, $B_\parallel$ is the component of the screen's magnetic field
along the line of sight in gauss, and
$s_{\rm pc}$ is the path length through the screen in parsecs. Any electron-proton
plasma inside the jet also causes rotation, although the magnitude of $\Delta\chi$
is reduced by the inverse of the average electron Lorentz factor $\langle \gamma \rangle$,
where $\gamma \equiv E/(m_{\rm e}c^2)$. Furthermore, when the rotation is distributed
through the emission region, $\Delta\chi$ is different for the different path lengths to
the emission sites along the line of sight. Because of this, the net effect is
partial depolarization of the radiation through cancellation of mutually orthogonal
components of the rotated polarization vectors from the different sites.

The linear polarization of any part of the source that is optically thick to
synchrotron self-absorption is reduced by a factor $\sim 7$ from the maximum value---to
$\sim 10\%$ in the uniform field case. Further reductions caused by a lower degree
of order of the magnetic field render self-absorbed synchrotron emission nearly
unpolarized in most cases. The EVPA is transverse to that of optically thin emission, and is
therefore parallel to the direction of the mean magnetic field.

Incoherent synchrotron radiation is circularly polarized at a level
$p_{\rm c} \sim \gamma^{-1}(\nu)$, where $\gamma(\nu)$ is the Lorentz factor 
of the electrons whose critical frequency is the
frequency of observation $\nu$: $\gamma(\nu) = 6.0\times 10^{-4}[\nu/(B\delta)]^{1/2}$,
where $B$ is the magnetic field strength. This is generally much less than 1\%.
The circular polarization observed in the compact jets, on the other hand, can be
as high as 1-3\% \cite{W98}. The likely cause is Faraday conversion of linear to circular
polarization by electrons (and positrons, if present) with $\gamma \sim 1-10$ \cite{JO77}.
\begin{figure}
\centering
\includegraphics[height=7.5cm]{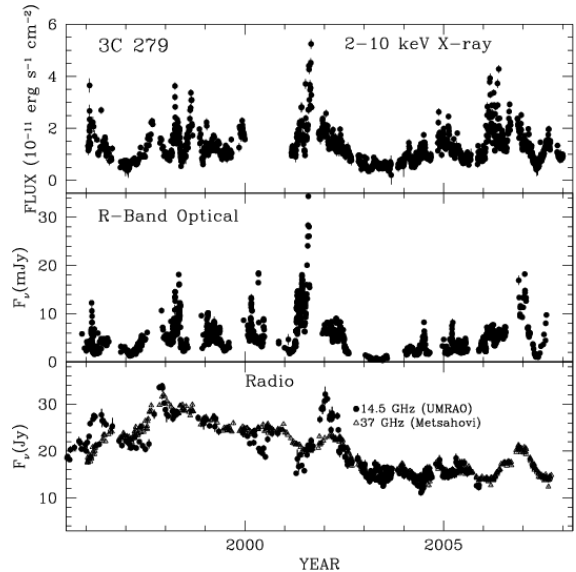}
\caption{Variation of flux vs. time of the quasar 3C~279\index{3C~279} at X-ray, optical, and radio
frequencies. From data of author and collaborators}
\label{fig:4}
\end{figure}
\subsection{Spectral Energy Distribution} 

A striking property of jets is
their ability to radiate profusely from radio to $\gamma$-ray frequencies. 
As Figs. \ref{fig:1} and \ref{fig:2} illustrate, jets can be prominent in optical and
X-ray images even at distances of hundreds of kiloparsecs from the nucleus. As seen in
the quasar 3C~279 (Figs. \ref{fig:3} \& \ref{fig:4} and \cite{Hart99}), the
compact, parsec-scale jets of prominent blazars imaged by VLBI
are the sites of bright radio through $\gamma$-ray emission that varies---often quite
dramatically---on time scales as short as hours or even minutes.
In some BL Lac objects, the fluctuating emission extends up to TeV $\gamma$-ray energies. 
The variations are often correlated across wavebands, sometimes with time delays.

\begin{figure}
\centering
\includegraphics[height=5.8cm]{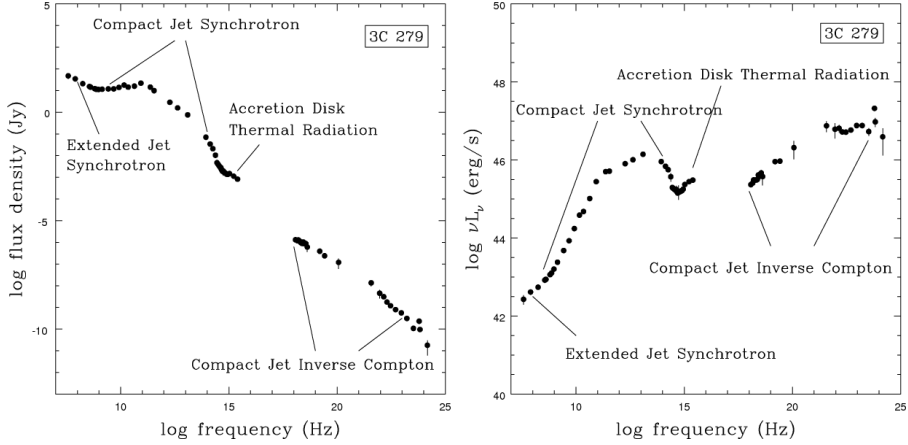}
\caption{{\bf (Left)} Continuum spectrum of the quasar 3C~279. {\bf (Right)} Same plot
but with spectral luminosity along the vertical axis. The main emission components are
indicated. From data compiled by the NASA Extragalactic Database plus data of the author
and collaborators}
\label{fig:5}
\end{figure}

Figure \ref{fig:5} displays the full electromagnetic continuum spectrum of 3C~279, with
the main physical components of the emission identified. Synchrotron emission from the
extended jet dominates at the lowest radio frequencies, while synchrotron radiation
from the compact jet supplies most of the flux from GHz to optical frequencies. Thermal
blackbody radiation from the accretion disk is prominent at ultraviolet frequencies,
while inverse Compton scattering by relativistic electrons in the compact jet produces
most of the X-rays at $> 1$ keV and $\gamma$ rays at $< 100$ GeV energies in this quasar.

The plot in the right panel of Fig. \ref{fig:5} emphasizes the spectral bands containing
most of the \emph{apparent} luminosity. Translation to \emph{actual} luminosity,
however, requires that we remove the effects of relativistic beaming by dividing by
$\delta^{2+\alpha}$ if the flux is steady and by $\delta^{3+\alpha}$ if the spectrum
corresponds to a temporary outburst. Since the Doppler factor of 3C~279 is
$\delta = 24\pm 6$ \cite{J05}, this requires division of the apparent luminosity of
the jet by a factor
exceeding 1000. In fact, the accretion disk is the most intrinsically luminous component
in most quasars. This does not, however, imply that the \emph{kinetic power} of jets
is a negligible fraction of the overall luminosity, since jets are rather
radiatively inefficient.

We can calculate the optically thin synchrotron flux density in the power-law
portion of the spectrum of the steady-state jet by adopting a conical geometry
with a magnetic field whose strength decreases with cross-sectional radius $R$,
$B = B_0(R/R_0)^{-b}$, and with a power-law electron density per unit energy,
$N(E) = N_0 (R/R_0)^{-a} E^{-(1+2\alpha)}$. Here $R_0$ is the radius
at the location where the emission turns on, presumably the point farthest upstream
where electrons are accelerated to energies high enough to radiate at the rest
frequency $\nu(1+z)/\delta$. Because of the gradients, we need to integrate
the emission coefficient \cite{Pac70} over volume, divide by $4\pi$ times the
square of the luminosity distance $D_l$, and apply appropriate relativistic
corrections:
\begin{equation}
F_\nu \approx (4\pi D_l^2)^{-1} c_1(\alpha) N_0 B_0^{1+\alpha} \delta^{2+\alpha}
(1+z)^{1-\alpha} \nu^{-\alpha} \phi^2 \int_{Z_0}^{\infty} Z^{2-a-(\alpha+1)b}dZ,
\label{eq:F}
\end{equation}
where, in cgs units, $c_1(\alpha) = 1.01\times 10^{-18}$, $3.54\times 10^{-14}$,
$1.44\times 10^{-9}$,
$6.30\times 10^{-5}$, and $1.5\times 10^{5}$ for $\alpha =$ 0.25, 0.5, 0.75, 1.0, and
1.5, respectively, incorporates numerical and physical constants (see \cite{JM04,M87}
for a tabulation), $z$ is the redshift of the host galaxy, $\phi$ is the opening
half-angle of the jet in radians, Z is distance in pc from the black hole along the jet
axis, and $Z_0 = R_0/\sin\;\phi$. The upper limit to the integral is
actually $Z_{\rm max}(\nu)$, the point where there are no longer substantial
numbers of electrons that can radiate at rest frequency $\nu(1+z)/\delta$. However,
if $a > 2$ and $b > 1$, we can use infinity
in practice as long as $Z_{\rm max}(\nu) \gg Z_0$. Equation (\ref{eq:F}) is valid under the
approximation that the Doppler factor is essentially the same across the jet, which
requires that $\Gamma$ does not depend on $R$ and that either the viewing angle
$\theta \gg \phi$ or $\phi \ll \Gamma^{-1}$. A more accurate calculation requires
numerical integration over polar and azimuthal angles inside the jet.

Note the flatness of the radio spectrum in the left panel of Fig. \ref{fig:5}, a
property caused by opacity
plus gradients in magnetic field strength and density of relativistic electrons.
We can derive the shape of the spectrum by considering the ideal case of a jet
pointing directly at us with an opening angle $\phi \ll \Gamma^{-1}$ and a
Doppler factor that does not vary significantly with $R$. We further approximate that
emission with optical depth $\tau > 1$ is completely opaque, while that at lower
optical depths is completely transparent. This means that all the emission comes
from radial distances $Z > Z_{\rm m}$, where $Z_{\rm m}$ corresponds to the
position in the jet where the optical depth along the line of sight is unity.
In order to derive this quantity, we again need to integrate:
\begin{equation}
\tau \sim 1 \approx c_2(\alpha) N_0 B_0^{0.5(2\alpha+3)}
[{\delta\over{\nu(1+z)}}]^{0.5(2\alpha+5)} \int_{Z_{\rm m}}^{\infty}
Z^{-a-0.5(2\alpha+3)b}dZ,
\end{equation}
where $c_2(\alpha)$ is $2.29\times 10^{12}$, $1.17\times 10^{17}$,
$6.42\times 10^{21}$, and $3.5\times 10^{26}$, for $\alpha =$ 0.25, 0.5, 0.75,
and 1.0, respectively. We can solve for $Z_{\rm m}$ after evaluating the integral:
\begin{eqnarray}
Z_{\rm m} \approx  &\big(c_2(\alpha) [a+0.5(2\alpha+3)b-1]^{-1} N_0
B_0^{0.5(2\alpha+3)} \times \nonumber\\
&[{\delta\over{\nu(1+z)}}]^{0.5(2\alpha+5)} \big)^{1/[a+0.5(2\alpha+3)b-1]}.
\label{eq:Zm}
\end{eqnarray}
We can now calculate the approximate flux at frequency $\nu < \nu_{\rm m}$, where
$\nu_{\rm m}$ is the frequency at which the source is transparent down to $Z_0$,
by replacing $Z_0$ by $Z_{\rm m}$ in the lower limit of the integral in
(\ref{eq:F}). (We can derive an expression for $\nu_{\rm m}$ by replacing $Z_{\rm m}$
with $Z_0$ in (\ref{eq:Zm}) and solving for $\nu$.) Since the main result we
want is the dependence on frequency, we express the
answer as a proportionality:
\begin{equation}
F_\nu(\nu < \nu_{\rm m}) \propto \nu^\zeta,
\end{equation}
where
\begin{equation}
\zeta = {{4\alpha(b-1) + 5(a+b-3)}\over{2(a-1)+(2\alpha+3)}}.
\end{equation}

A completely flat spectrum, with $\zeta = 0$, is possible if $a=2$ and $b=1$, as
proposed by K\"onigl \cite{K81}. While this value for $b$ corresponds to the
expected dependence of a transverse magnetic field on $R$, the value $a=2$
would only hold if adiabatic expansion cooling did not occur. If, on the
other hand, expansion cooling is the main energy loss mechanism for the electrons,
we expect $a = (4/3)\alpha + 2$ in the case of a jet with constant flow Lorentz factor\index{Lorentz
factor} $\Gamma$. Since actual jets usually contain a number of knots in addition
to the core, the flat spectrum must be caused by a superposition of the individual
spectra of the various features, each peaking at different frequencies. The
fact that these tend to produce composite spectra that are, to a rough approximation,
flat, implies that K\"onigl's ansatz is globally correct, even if it might not
apply locally. For example, standing and/or moving shock waves in the jet can
convert flow energy into plasma energy, whereas adiabatic expansion cooling does
the reverse. As long as only an insignificant fraction of the total flow energy is
lost to radiation, the
overall spectrum will follow the $a=2$, $b=1$ case. In fact, careful analysis of
the jet in 3C~120 over several orders of magnitude in length scale indicates that it
maintains the values $a=2$ and $b=1$ \cite{WBU87}.

\section{Physical Processes in AGN Jets}
\label{sec:3}
Considerable progress has been made in the interpretation of observations of jets
by assuming that the main physical processes governing jets are those of gas
dynamics and magneto-hydrodynamics (MHD). Plasma physics in the relativistic regime
must also play a role in the heating and motions of the particles that compose jets.
Unfortunately, our understanding of the relevant plasma phenomena is hindered by
insufficient observations coupled with the likely complexity involved when the
mean particle energy is relativistic. In addition, electrodynamics might govern the
dynamics close to the black hole, and electrical currents can play an important
role there and farther downstream. Here we will discuss some of the physical
processes that have been successfully applied to explain a wide range of
observations of jets.

\subsection{Launching of Jets}
\label{sec:launch}
The basic requirement for producing a supersonic, highly collimated flow is a confinement
mechanism coupled with a strong negative outward pressure gradient. In principle,
this is possible with gas dynamics if the pressure of the interstellar medium in
the nucleus drops by many orders of magnitude from the central regions to the
point where the jet reaches its asymptotic Lorentz factor\index{Lorentz factor} \cite{BR74}. However,
the most extreme jets are accelerated to $\Gamma > 10$ and collimated to
$\phi < 1^\circ$ within $\sim 1$ pc from the central engine \cite{J05}. At this
location, the pressure in the jet is typically of order 0.1 dyne cm$^{-2}$
(e.g., \cite{M07}). Gas dynamical acceleration and collimation would then
require an external pressure $\sim 10^4$ dyne cm$^{-2}$ within $\sim 0.01$ pc
of the black hole. If this were provided, for example, by a hot ($\sim 10^8$ K)
wind emanating from the accretion disk, the X-ray luminosity would be enormous,
$\sim 10^{50}$ erg s$^{-1}$.

For this reason, magnetic launching is considered to be the
driving force behind most relativistic jets in AGN. The differential rotation of the
ergosphere of a spinning black hole and/or accretion disk causes the component
of the magnetic field lines of the accreting matter to wind up into a helix. The
toroidal component of the wound-up field, $B_{\rm tor}$, provides a pinching
force through the hoop stress, with magnitude
$F_{\rm pinch} = B^2_{\rm tor}/(4\pi R)$, directed toward the jet axis.
The magnetic field expands with distance from the black hole, lowering the
magnetic pressure. This creates a strong pressure gradient along the axis, which
accelerates the flow. In order for the velocity to become highly relativistic,
the magnetic energy density must greatly exceed the rest-mass energy density
at the base of the jet. Hence, the flow is initially dominated by Poynting flux
that is transferred to kinetic flux of the particles in the jet toward
greater axial distances $Z$.

The magnetic acceleration essentially stops when the kinetic energy density of the
particles reaches equipartition with the magnetic energy density. Therefore, the
final flow Lorentz factor is \cite{VK04}
\begin{equation}
\Gamma_{\rm max}(Z) \sim (B^2_{\rm i})/[16\pi n(Z)\langle m\rangle c^2],
\end{equation}
where $n(Z)\langle m\rangle$ is the mass density of particles at position
$Z$, and $B_{\rm i}$ is the magnetic field at the base of the jet.

In the magnetic (as well as gas dynamical) acceleration model, the opening angle
$\phi$ is inversely proportional to the final Lorentz factor, as observed \cite{J05}.
The gradual acceleration implies that any radiation from the innermost jet is weakly
beamed relative to that on parsec scales. Therefore, one should expect that most of
the observed emission arises from regions near and downstream of the core
seen on millimeter-wave VLBI images, even at high frequencies where the entire jet
is transparent.

\subsection{Gas Dynamics of Jets}
\label{sec:HD}
Since jets are
long-lived phenomena, we can assume that, to first order, they represent steady
flows. Then the relativistic Bernoulli equation\index{Bernoulli equation} applies:
\begin{equation}
\Gamma (u+p)/n = constant,
\label{eq:Gam}
\end{equation}
where $u$ is the energy density of the plasma, $p$ is the pressure, and
$n$ is the number density of particles, all measured in the co-moving reference
frame. If the mean energy per particle $\langle E \rangle$ greatly exceeds
the mean rest-mass energy,then
$u = 3p = n\langle \gamma m\rangle c^2$. In addition, the ideal gas law for a relativistic plasma can be expressed as $p \propto n^{4/3}$.
We can use these expressions in (\ref{eq:Gam}) to obtain
\begin{equation}
\Gamma(Z) = \Gamma(Z_0)[p(Z_0)/p(Z)]^{1/4},
\label{eq:Gam2}
\end{equation}
where $Z_0$ is a reference point. This is the relativistic version of
Bernoulli's principle, which states that the velocity of a flow is inversely
related to its pressure.

We could also rearrange
the ideal gas law to write $\langle \gamma\rangle \propto n^{1/3}$ and
\begin{equation}
\Gamma(Z) = \Gamma(Z_0)\langle \gamma(Z_0)\rangle/\langle \gamma(Z)\rangle.
\label{eq:Gam3}
\end{equation}
This formulation
emphasizes that the acceleration of the flow results from conversion
of internal energy into bulk kinetic energy. Significant increase in the
flow speed through gas dynamics is no longer possible if
$\langle \gamma(Z)\rangle$ is not substantially greater than unity. Beyond
this point, the jet will be ballistic, coasting at a constant velocity until
it is disturbed.

The cross-sectional area of a jet will tend to expand and contract such that
the pressure at the boundary matches the external pressure. The maximum opening
angle $\phi$, however, is the inverse of the Mach number at the point where the 
pressure drop occurs, $\phi_{\rm max} = v_{\rm s}/(\Gamma\beta c)$, where
$v_{\rm s}$ is the local sound speed.

Since the interstellar medium is dynamic,
and since the pressure in the jet is subject to time variability, a jet is
likely to be subject to mismatches in pressure at the boundary. If modest in
magnitude, the difference in pressure is accommodated through sound waves that
communicate the pressure imbalance to the jet interior. This causes both
oscillations in the cross-sectional radius as well as compressions and rarefactions
inside the jet \cite{CF48}. The Lorentz factor is higher in the lower pressure
regions and lower where the pressure is greater. If the pressure mismatch exceeds
$\sim 50$\%, oblique ``recollimation'' shock waves form to adjust the flow
more abruptly. If the jet is sufficiently circularly symmetric, the shock waves will
be conical. If the symmetry is nearly perfect, the conical shock is truncated by
a strong shock aligned perpendicular to the axis, called a ``Mach disk\index{Mach disk}.''
Since the role of a shock front is to decelerate a disturbed supersonic flow
to subsonic velocities, the flow will cease to be highly relativistic until it
reaccelerates where it encounters a pronounced rarefaction downstream of the Mach
disk. Shock waves set up by pressure mismatches with the surrounding medium
are often labeled ``external'' since they are driven by influences from outside
the jet.

``Internal'' shock waves can result from changes in the jet velocity or injected
energy at or near the base. A major disturbance is required to generate a shock
when the flow velocity is near the speed of light, since the relative velocity
needs to be supersonic. If the local sound speed is the
fully relativistic value, $c/\sqrt{3}$, then the Lorentz factor of the faster flow
must be at least a factor $\sim 2$ higher than that of the slower flow that it
overtakes.

A shock compresses the plasma so that the density rises by a factor
$\xi$ across the shock front. The magnetic field component that
lies parallel to the shock front increases by the same factor, while the component
transverse to this direction remains unchanged. The maximum value of $\xi$ is 4
in a thermal plasma; it can be higher in relativistic plasma, but if the jet flow
is already highly relativistic, disturbances are unlikely to be sufficiently strong
relative to the quiescent flow for values of $\xi$ greater than about 2 to be common.
In the reference frame of the shock front, the undisturbed plasma flows in at
a supersonic velocity and exits at a subsonic speed. Unless the sound speed is
close to $c$---the maximum value of $c/\sqrt{3}$ requires that the plasma be highly 
relativistic, with $\langle \gamma\rangle \gg 1$---this means that the shocked plasma
will have a bulk Lorentz factor only slightly lower than that of the shock front.

\subsection{Magnetohydrodynamics of AGN Jets}
\label{sec:MHD}
The gas dynamical description of a jet is valid as long as the energy density
of the magnetic field is significantly less than that of the particles.
In this case, the field is ``frozen in'' and will follow the plasma. Close to
the black hole, however, the magnetic field plays a major role in the dynamics,
and even on kiloparsec scales the field energy is probably roughly in
equipartition with that of the particles.

Section \ref{sec:launch} above briefly
summarizes the MHD forces that can accelerate and collimate a jet in the inner
parsec. This depends on a strong toroidal component of the magnetic field,
which itself requires a radial electric current. Since the system has considerable
angular momentum, the streamlines that the flow follows trace out a spiral pattern
about the axis \cite{V06}.

When the magnetic field is mainly toroidal, kinks that occur in the jet
are subject to a current-driven instability that causes the kinks to grow
\cite{Beg98}. Simulations indicate that the net result is to convert Poynting
flux into internal energy of the plasma \cite{H06}.

We therefore have a crude theoretical description of a jet that is collimated by
a toroidal magnetic field and accelerated by a decreasing magnetic pressure
gradient, but eventually subject to an internal instability that heats the
plasma. By the time such a jet reaches parsec scales, it has a relativistic
flow velocity, contains a magnetic field that is at least somewhat chaotic,
and is loaded with relativistic electrons (and either protons or positrons, or both).
These are the necessary ingredients to produce the jets observed on VLBI images,
with bright synchrotron and inverse Compton emission and fairly weak linear
polarization.
\subsection{Instabilities in Jets}
\label{sec:instab}
\begin{figure}
\centering
\includegraphics[height=7.5cm]{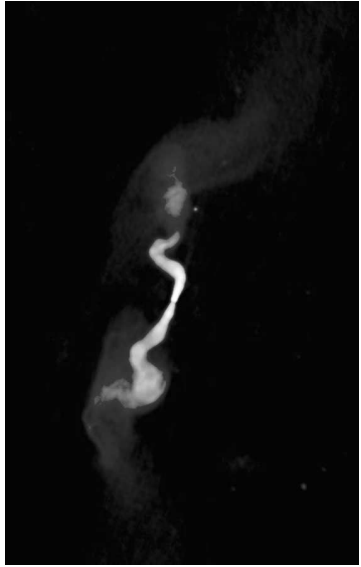}
\caption{Radio image of the FR~I radio galaxy 3C~31\index{3C~31} showing a two-side jet that becomes
globally unstable. The nucleus is in the middle of the straight section of the jet.
Image courtesy of NRAO/AUI/NSF, data published in \cite{LB02}}
\label{fig:6}
\end{figure}

As the parsec to kiloparsec-scale continuity of jets like the one in 3C~273\index{3C~273}
displayed in Fig. \ref{fig:1} demonstrates, many jets are stable over
extremely large distances.
Jets that are completely free, i.e., containing higher pressure than their 
surroundings and propagating ballistically, are inherently stable. However, if 
their cross-sections are able to adjust to maintain pressure equilibrium with the 
external medium, they are subject to the Kelvin-Helmholtz instability\index{Kelvin-Helmholtz instability} that occurs
when two fluids have different velocities adjacent to a tangential discontinuity.
These instabilities have numerous modes that are divided into two classes, body
modes and surface modes. The body modes cause departure of the cross-sectional
geometry from circular symmetry as well as oscillations of the
transverse radius and filamentary structure \cite{H06}. Figure \ref{fig:6}
shows an example of such a jet. The surface modes
generate turbulence and velocity shear at the boundaries that can permeate the 
entire jet cross-section as the instability grows.

The reason why jets can remain largely intact hundreds of kiloparsecs from
their site of origin in the nucleus is that Kelvin-Helmoltz instabilities
are suppressed at higher Mach numbers and Lorentz factors, and when there
is a strong axial magnetic field. These characteristics are found in
the most powerful jets, those of the FR II quasars and radio galaxies.

Since they can lead to the development of turbulence and formation of shocks,
instabilities might be the cause of much of the emission seen on parsec
and kiloparsec scales \cite{Sik05}. In addition, much of the rich structure
seen in some jets might owe its existence to such instabilities.
The turbulence in the boundary region also promotes entrainment of material from
the surrounding medium \cite{Bick94,Laing96}, a process that slows the jet down 
\cite{Laing99}, eventually leading to its disruption in FR~I sources.
\subsection{Cross-Jet Structure}
\label{sec:cj}
Jets can possess structure transverse to the axis whether or not instabilities
develop. In fact, theoretical models and simulations of
jet launching generally predict such structure
\cite{SPA89,Pun96,MKU00,VK04,HK06,McK06}. Although there can be a smooth gradient
in density and velocity \cite{VK04}, transverse structure is often described
as an ultra-fast spine surrounded by a more slowly moving sheath. The
spine is predicted by many theories to be composed
of an electron-positron pair plasma, with the sheath containing a normal
electron-proton mixture \cite{SPA89,Pun96,McK06}. While there is considerable
observational evidence for cross-jet structure \cite{Laing99,PGVY05,J05}, it is not yet
clear whether there is a smooth gradient in properties or a two-layer structure.

\subsection{Magnetic Field}
\label{sec:mag}
As discussed previously, we expect the magnetic field to have a tight helical
geometry close to the black hole. At the end of this acceleration and collimation
zone, current-driven instabilities mix the field into a more chaotic
configuration. Velocity shear can either align the magnetic field parallel
to the axis or generate turbulence that disorders the field. In the former case,
one can imagine closed loops of field lines that are stretched
by the relative velocity between faster flow closer to the axis and slower
layers closer to the boundary.

As in any spherical or conical flow in which magnetic flux is conserved, the axial 
component of
the magnetic field decreases as $R^{-2}$. The component transverse to the axis
should decrease as $R^{-1}$. There are numerous examples of extended jets in which
the expected transition from longitudinal to transverse magnetic field is evident
in the linear polarization \cite{Bridle84}. However, comparison of simulated
and observed polarization indicates that, while the toroidal to longitudinal
magnetic field ratio does increase with distance down the jet, the gradient in the
ratio is not as steep as predicted in the case of a magnetic field that is
frozen into the plasma \cite{LCB06}. There is instead a substantial component
of disordered field that has a more complex behavior.

\subsection{Energization of Electrons}

Since the radiation from AGN jets comes from electrons (and positrons, if any),
the sites of strong emission are strongly related to the processes
that accelerate electrons to Lorentz factors $\gamma$ that can exceed $10^6$
on both subparsec and kiloparsec scales in some jets. The power-law shape with
slope $-\alpha$ of
the optically thin synchrotron spectrum over $\sim 4$ orders of magnitude
(see Fig. \ref{fig:5}) implies that the electron energy distribution is also a
power law, with slope (on a log-log plot) $-(2\alpha+1)$ over at least 2 orders
of magnitude in $\gamma$.

A long-standing explanation for a power-law electron energy distribution is the
Fermi mechanism\index{Fermi mechanism} \cite{F49}. Particles repeatedly reflect off regions of
higher-than-average magnetic fields, gaining energy if the region was
moving toward the particle (in the plasma's rest frame)
and losing energy if it was receding. The net result is a power-law energy
distribution. If a particle gains an average energy
$\Delta E = \xi E$ per reflection and if the probability of staying in the acceleration
zone after a reflection is $\eta$, $\alpha = {\rm ln}\;\xi/(2{\rm ln}\;\eta)$.
The process can produce very high energies if $\eta$ is close to unity so that a
typical particle encounters many reflections before exiting the zone.

One site where the Fermi mechanism can be particularly efficient is at a shock
front. In this case, $\xi = (4\Delta u)/(3c)$, where $\Delta u$
is the increase in flow velocity across the shock front \cite{Bell78}.
Plasma waves are likely to be present on both sides
of the shock. These reflect the particles so that they pass back and forth across
the shock multiple times. Less systematic acceleration of particles can occur in
a turbulent region since, statistically, the number of approaching
collisions with cells of higher than average magnetic field
is greater than the number of receding collisions. The mean fractional
energy gain per reflection is $\xi \approx 4(\Gamma_{cell}v_{\rm cell}/c)^2$.
Therefore, the process is efficient if the turbulence is relativistic, as one might
expect if the jet has substantial transverse velocity gradients.

An unresolved issue with the Fermi mechanism is how it can produce a similar
power-law slope in remotely separate physical regions. For example, the
spectral index in the jet of 3C~120 $\alpha \approx 0.65$ over several orders
of magnitude in size scale \cite{WBU87}. If the bright emission regions are caused by
shocks, all the shocks would need to have the same compression ratio, which
would seem difficult to engineer. If turbulent processes energize the particles,
the physical conditions would again need to be more similar than one might expect
over such a large range of size scales.

\section{Physical Description of Features Observed in Jets}
\label{sec:physfeatures}
\subsection{The Core}
\label{sec:core}
The core is generally the most compact part of a jet that we can image at any
given frequency. This implies that an understanding of its nature can provide
insights into the physical structure of jets at and perhaps upstream of the
core's position. As we have discussed above, what is seen as the core on VLBI
images at radio frequencies might be the transition region between optically thick
and thin emission. However, it could also be a bright, stationary
emission structure a short distance downstream of that region.
If the core is the surface where the optical depth is near unity, its position
$Z_{\rm m}$ should move toward $Z_0$ and therefore toward the central engine at higher 
frequencies. This effect is apparent in some compact extragalactic jets
\cite{L98}, but not in others \cite{MPWBB06}. The implication
of the latter negative result is that the jet is not just a smooth flow punctuated
by superluminal knots, but also contains multiple bright stationary features.
In the quasar 0420$-$014, observations of fluctuating polarization in the core and at
optical wavelengths have been interpreted as the signature of turbulent jet plasma
passing through a standing conical shock system---perhaps an example of a recollimation
shock discussed in Sect. \ref{sec:HD}---that represents the physical
structure of the core \cite{Darc07}.

In some objects the core could be a bend in the jet where it becomes more closely
aligned with the line of sight. This would cause
an increase in the Doppler beaming factor relative to points farther upstream. This
cannot be the main reason for the appearance of cores, however, since it is less
likely for a jet to curve toward than away from the line of sight.

While jets are often approximately self-similar on parsec scales, this must break
down close to the central
engine in the flow acceleration and collimation zone (ACZ). The point where the
change must occur is indicated by the frequency above which the synchrotron spectrum
steepens, signaling that the jet emission is optically thin. According to
observations \cite{IN88}, this is in the range of tens to $> 1000$ GHz.
(It is $10^{11}$ Hz in Fig. \ref{fig:5}.)
The angular width of the jet at this point can be estimated by the value at
43 GHz, $\sim 50$ $\mu$arcsec times 43 GHz/$\nu_{\rm m}$. For a quasar with
redshift of order 0.5, this translates to $\sim 0.3$(43 GHz/$\nu_{\rm m}$) pc,
which is small (a few $10^{16}$ cm for $\nu_{\rm m} \sim 1000$ GHz) but still hundreds
of Schwarzschild radii even for a black hole mass $\sim 10^9$ M$_\odot$. The distance
of the end of the ACZ from the black hole might be many times larger than this.
The flux of radiation from a blazar jet inside the ACZ probably decreases toward
the black hole, since the relativistic beaming is weaker owing to the progressively
lower flow Lorentz factor \cite{M80}. In the jet of a radio galaxy viewed nearly
side-on, however, the emission should become stronger (as long as there are
energetic electrons) toward smaller distances from
the black hole, since the radiation is anti-beamed. This implies that
there should be a very short gap between the start of the jet and the beginning
of the counterjet, except for a limited section obscured by free-free opacity
in the accretion disk and dusty torus beyond the disk.
Furthermore, the flow should start out fairly broad and become more collimated
as it accelerates downstream. VLBA observations of the radio galaxy M87, whose
relatively local distance allows a linear resolution of
tens of gravitational radii, indicate that the flow is indeed quite broad
near the central engine \cite{JBL99}.

\subsection{Quasi-stationary Features}
\label{sec:hot}
Besides the core, there are often other bright features in blazar jets that are either
stationary or move at subluminal apparent speeds \cite{J01,K04,J05}. In straight jets,
these can be produced by recollimation shocks or instabilities (see Sect. \ref{sec:HD}
and Sect. \ref{sec:instab})
that compress the flow or excite turbulence. This should lead to particle acceleration
and amplification of the magnetic field, and therefore enhanced emission.

Bends can also cause quasi-stationary hotspots, either because the jet turns more into
the line of sight \cite{Alb93} or from the formation of a shock. An example of the latter
is the collision of the jet with an interstellar cloud. In this case, an oblique shock
forms to deflect the jet flow away from the cloud.
\begin{figure}
\centering
\includegraphics[height=5.5cm]{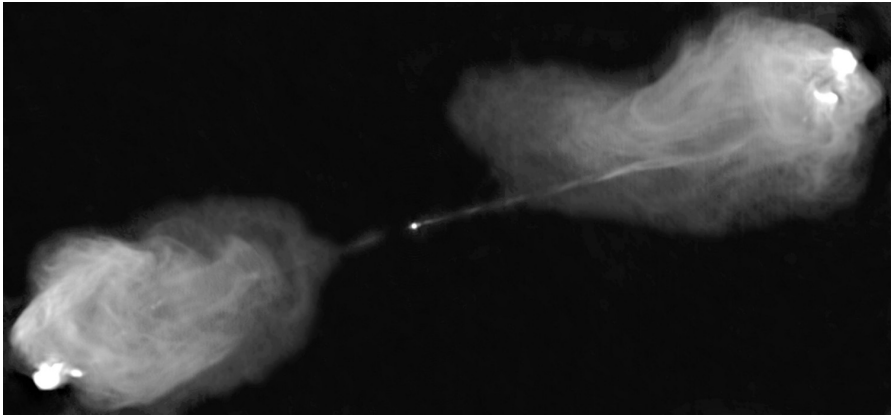}
\caption{Radio image of the FR~II radio galaxy Cygnus~A showing the small, bright core
at the center, two-side jet, billowy lobes, and hotspots at the outer edges of the lobes.
The western side approaches us, hence the lobe is farther from the core since the hotspots
and lobes move outward with time and we see the eastern lobe at an earlier time owing to
light-travel delay. Image courtesy of NRAO/AUI/NSF, published in \cite{PDC84}}
\label{fig:7}
\end{figure}

Long-lived jets propagate through their host galaxy and into the intergalactic medium.
In the case of low-luminosity, FR~I sources (e.g., the one shown in Fig. \ref{fig:6}), 
instabilities at the boundary and entrainment of the external gas decelerate the jet to
subsonic speeds and completely disrupt it. The more powerful FR~II jets generally survive
out to the point where a strong termination shock decelerates them to subsonic speeds. The
shock appears as a bright, compact hotspot, while the jet plasma spreads into a giant
turbulent ``lobe'' that emits radio synchrotron radiation (see Fig. \ref{fig:7}).
As inferred by the ratio of distances from the nucleus of the near and far lobes,
the termination shock moves subluminally away from the galaxy as the momentum of the jet
pushes away the external gas.

\subsection{Models for Superluminal Knots in Jets}
\label{sec:knots}
The bright knots that appear to move superluminally down compact jets (see Fig.
\ref{fig:3}) must represent
regions of higher relativistic electron density and/or magnetic field than the ambient
jet. They could be coherent structures, such as propagating shock waves caused by
surges in the energy density or velocity injected into the jet at its origin
near the central engine. However, they could also be ``blobs'' of turbulent plasma
with electrons accelerated by the second-order, statistical Fermi process, i.e., random 
encounters with magnetic irregularities moving in various directions. Or, they might
be ribbons or filaments generated by instabilities that move down the jet. It is likely
that each of these possibilities occurs in some jets.
\begin{figure}
\centering
\includegraphics[height=10cm]{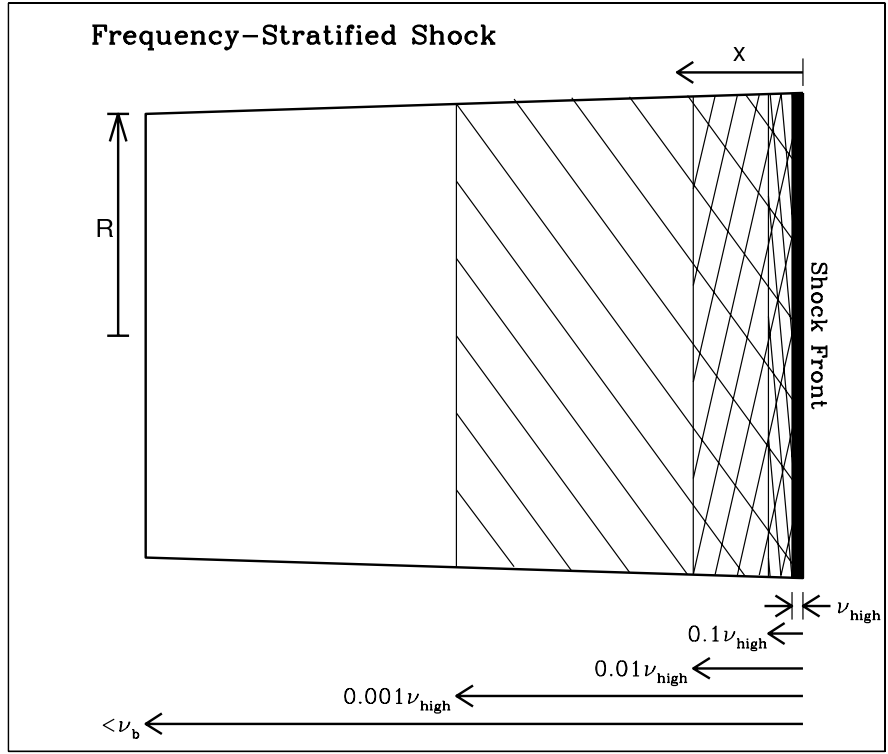}
\caption{Frequency stratification in a square-wave shock, drawn to scale. The electrons
are accelerated at the shock front and drift behind the shock while losing energy to
radiative losses. This example is for a forward propagating shock; the same
principle is valid for a reverse or stationary shock, as well as for a shock that
is oblique to the jet axis. See \cite{MG85} for more details}
\label{fig:8}
\end{figure}

Shock waves are the most common interpretation of superluminal knots, since basic
shock models have enjoyed considerable success at reproducing time variations 
of continuum spectra and polarization \cite{MG85,HAA85,HAA89}. At frequencies above
$\sim 10$ GHz, radiative energy losses of the electrons can be important. In this
case, if we assume that electrons are accelerated only at the shock front, they
will lose their ability to produce synchrotron radiation at high frequencies as they
advect away from the
front and lose energy, since they cannot radiate much above their critical frequency,
$\nu_{\rm crit} = 2.8\times 10^6 B\gamma^2$ (in cgs units). This results in frequency 
stratification,
with the highest frequency emission confined to a thin layer behind the shock front
and progressively lower frequency emission occupying larger volumes. Figure
\ref{fig:8} illustrates the effect. We can derive a
proportionality between distance $x$ behind the front and the frequency of emission
that is limited to a shell of thickness $x$ if we assume that
the magnetic field and advection speed $v_{\rm ad}$ are constant out to distance
$x$ from the shock front:
\begin{equation}
x(\nu) \propto \nu^{-1/2}.
\end{equation}
Here, we have applied the solution to the synchrotron energy loss equation,
$\gamma_{\rm max} \approx (1.3\times 10^{-9}B^2x/v_{\rm ad})^{-1}$. The volume of the 
emitting region is then $\pi R^2x \propto \nu^{-1/2}$. This steepens the optically thin
spectral index from $\alpha$ to $\alpha+0.5$ at frequencies $\nu \gg \nu_{\rm b}$,
where $\nu_{\rm b}$ is the frequency at which $x(\nu)$ equals the entire length of
the shocked region, and $-\alpha$ is
the spectral slope of the emission coefficient. It also affects the time evolution
of the flux density and turnover frequency as the shock expands while it propagates
down the jet. In particular, at frequencies where synchrotron losses are greater than
expansion cooling, the peak in the spectrum propagates toward lower frequencies while
the flux density at the peak remains roughly constant \cite{MG85}. This can contribute
to the flatness of the overall synchrotron spectrum at radio frequencies.

The stratification results in time lags in the optically thin synchrotron radiation
during the outburst caused by the formation of the shock, with higher frequency
variations leading those at lower frequencies. Even beyond the point where the magnetic
field is too low for synchrotron losses to compete with expansion cooling, time
delays at radio frequencies continue owing to the outburst becoming transparent at
progressively lower frequencies with time.
Lower amplitude flares and flickering (cf. the light curves displayed in Fig. \ref{fig:4})
can be explained as the consequence of shocks compressing turbulent cells that they
pass \cite{MGT92}. In this case, the higher frequency variations in
flux density and polarization occur on shorter timescales because of the smaller
volumes involved.

For the sake of simplicity, shocks in jets are usually modeled as
transverse square waves in attempts to reproduce the observed variability of emission
and polarization. Such shocks will compress the magnetic field component lying
parallel to the shock front. In most jets showing superluminal motion, the
orientation of the shock front is aberrated to appear nearly transverse to
the jet axis in the observer's frame. This should yield linear polarization
with electric vectors aligned along the jet axis. A large sample of quasar jets
\cite{LH05} observed with VLBI possess a distribution of polarization directions
that is inconsistent with this assumption. Instead, the magnetic field seems to be
mostly turbulent, with a small uniform component that is parallel to the jet axis.
Other polarization observations \cite{AAH03} indicate that, if most superluminal knots
are shocks, the fronts must be oblique to the jet axis in most cases.

The main differences between shock and blob models are the absence of frequency
stratification in the latter and that it is unnecessary for the particles and
magnetic field in a blob to be directly related to those of the surrounding
jet plasma. In the case of an outburst of 3C~273\index{3C~273}, an attempt to model the event in terms
of an expanding blob led to the nonsensical requirement that the magnetic field would
need to have decreased while the particle density increased to reproduce the
time behavior of the flare's continuum spectrum \cite{MG85}. On the other hand, it
is common to model X-ray synchrotron and $\gamma$-ray inverse Compton flares in
BL Lac objects as blobs (e.g., \cite{Kat99}).

\subsection{Internal Energy and Kinetic Power}
\label{sec:E}

There are basically two methods for determining the energy contained in relativistic
particles in extragalactic jets. The first is to assume equipartition between the
magnetic and particle energy densities, which is similar to calculating the minimum
energy required to produce the synchrotron emission \cite{B59}. We can express the
ratio of relativistic particle to magnetic energy density as
$u_{\rm p}/u_{\rm mag} \equiv A$, with $A = 1$
corresponding to equipartition. If $k$ is the fraction of particle energy that is
contained in neither electrons nor positrons, the energy density in particles is
\begin{equation}
u_{\rm p} = (1+k)N_0 \int^{\gamma_{\rm max}m_{\rm e}c^2}_{\gamma_{\rm min}m_{\rm e}c^2}
E^{-2\alpha} dE,
\label{eq:up}
\end{equation}
which we can set equal to $Au_{\rm mag} = AB^2/(8\pi)$.

If we consider a section of volume $V$ inside the jet, with radiation coming from
some fraction $f$ (the ``filling factor'') of that volume, and observe it to have a flux
density $F_\nu$ at a frequency $\nu$ where the emission is optically thin, we can write
\begin{equation}
F_\nu \approx c_1(\alpha) D_l^{-2} fV N_0 B^{1+\alpha} \delta^{3+\alpha}
(1+z)^{1-\alpha} \nu^{-\alpha}.
\label{eq:FE}
\end{equation}
Here we have assumed that the measurement is for a moving knot; if it is a section
of the steady-state jet, the exponent of $\delta$ should be $2+\alpha$. 
We can solve this equation for $N_0$, substitute the result for $N_0$ in
(\ref{eq:FE}), and set $u_{\rm p}$ equal to $AB^2/(8\pi)$. After evaluation
of the integral, some algebraic manipulation, and conversion of various
parameters into more readily observable quantities, we arrive at
\begin{eqnarray}
B = &c_4(\alpha) \delta^{-1} (1+z)^{(5+\alpha)/(3+\alpha)}~
\times~~~~~~~~~~~~~~~~~~~~~~~~~~~~&\nonumber\\
&\times \big[{{1+k}\over{Af}} g(\alpha,\gamma_{\rm min},\gamma_{\rm max})D_{\rm Gpc}^{-1}
\Theta_{\rm as}^{-3} F_{\rm mJy} \nu_{\rm GHz}^{\alpha}\big]^{1/(3+\alpha)} \mu{\rm G}.
\label{eq:B}
\end{eqnarray}
Here $D_{\rm Gpc}$ is the luminosity distance in Gpc, $\Theta_{\rm as}$ is the
angular size in arcseconds (under the approximation of spherical symmetry),
$F_{\rm mJy}$ is the flux density in milliJanskys (1 mJy = $10^{-26}$ erg s$^{-1}$
cm$^{-2}$ Hz$^{-1}$), $\nu_{\rm GHz}$ is the observed frequency in GHz, and
\begin{equation}
g(\alpha,\gamma_{\rm min},\gamma_{\rm max}) \equiv (2\alpha - 1)^{-1}
[\gamma_{\rm min}^{-(2\alpha-1)} -
\gamma_{\rm max}^{-(2\alpha-1)}]
\label{eq:G}
\end{equation}
except for the case $\alpha = 0.5$ for which
$g = {\rm ln}(\gamma_{\rm max}/\gamma_{\rm min})$.
The function $c_4(\alpha)$, which is tabulated in \cite{JM04}, has a value of
1.9, 7.8, 25, 74, 170, and 380 for $\alpha$ = 0.25, 0.5, 0.75, 1, 1.25, and 1.5,
respectively. The total energy density is $(A+1)B^2/(8\pi)$, hence the kinetic
power is $(\pi R^2 c) \Gamma^2 (A+1)B^2/(8\pi)$.

For typical spectral indices in extended jets, $\alpha = 0.8\pm{0.2}$, it is mainly
$\gamma_{\rm min}$ that determines the value of the function $g$.
Unfortunately, we cannot in general determine $\gamma_{\rm min}$ accurately, since
electrons at this energy radiate mainly in the MHz range or lower where the
emission has many components that are difficult to disentangle. We also do not
have reliable methods to estimate the parameters $f$, $A$, or $k$.
The usual procedure is to assume $f=1$, $A=1$ and $k=1$, and some fairly
high value of $\gamma_{\rm min}$, e.g., 100--1000, in order to derive a lower
limit to the kinetic power. The powers thus calculated can be very high in
quasar jets, exceeding $10^{46}$ erg s$^{-1}$ in many cases, and as high as
$10^{48}$ erg s$^{-1}$ if $\gamma_{\rm min} \sim 10$ or if
the ratio of proton to electron energy density $k$ is closer to 100, as is the
case for cosmic rays in our Galaxy.
The observational data are usually consistent with rough equipartition between
the energy densities in electrons and in magnetic field under the assumption that
$k \sim 1$.

The second method for deriving the energy content of jets is to analyze compact knots
that are synchrotron self-absorbed, in which case there are two equations, one
for the flux density and the other for the optical depth (which is close to unity at
the frequency $\nu_{\rm m}$ of maximum flux density). We can then solve separately for
$B$ and $N_0$ (see, e.g., \cite{M83}). The magnetic field can be expressed as
\begin{eqnarray}
B &=& 3.6\times 10^{-5}c_{\rm b}(\alpha) \Theta_{\rm mas}^4 \nu_{\rm m,GHz}^5
F_{\rm m,Jy}^{-2} \delta (1+z)^{-1}\nonumber\\
&=& 0.011c_{\rm b}(\alpha) T_{\rm m,11}^{-2} \nu_{\rm m,GHz} \delta (1+z)^{-1},
\label{eq:umag}
\end{eqnarray}
where the units are indicated by the subscripts and $T_{\rm m,11}$ is the brightness
temperature at $\nu_{\rm m}$ in units of $10^{11}$ K. The function $c_{\rm b}(\alpha)$
is 0.50, 0.89, 1.0, and 1.06 for $\alpha =$ 0.25, 0.5, 0.75, and 1.0, respectively.
The similar equation for $N_0$ is
\begin{eqnarray}
N_0 &=& 0.012c_{\rm n}(\alpha) D_{\rm Gpc}^{-1} \Theta_{\rm mas}^{-(7+4\alpha)}
\nu_{\rm m,GHz}^{-(5+4\alpha)} F_{\rm m,Jy}^{3+2\alpha} \delta^{-2(2+\alpha)}
(1+z)^{2(3+\alpha)}\nonumber\\
&=& 0.012(17.7)^{-(3+2\alpha)})c_{\rm n}(\alpha) D_{\rm Gpc}^{-1} T_{\rm m,11}^{3+2\alpha}
\nu_{\rm m,GHz} \Theta_{\rm mas}^{-1} (1+z)^3 \delta^{-1},
\label{eq:N0}
\end{eqnarray}
where $c_{\rm n}(\alpha)$ is 660, 22, 1.0, and 0.049 for $\alpha =$ 0.25, 0.5, 0.75,
and 1.0, respectively.

We can therefore express the ratio of the relativistic electron to magnetic
energy density mainly in terms of the brightness temperature:
\begin{equation}
u_{\rm re}/u_{\rm mag} \approx g_{\rm u}(\alpha,\gamma_{\rm min},\gamma_{\rm max})
{T_{\rm m,11}}^{7+2\alpha} (D_{\rm Gpc} \nu_{\rm m,GHz} \Theta_{\rm mas})^{-1}
(1+z)^5 \delta^{-3},
\label{eq:urat}
\end{equation}
where
\begin{eqnarray}
g_{\rm u}(\alpha,\gamma_{\rm min},\gamma_{\rm max}) \approx &0.48\gamma_{\rm max}^{0.5}
~~~~~~~~~~~~&(\alpha=0.25)\nonumber\\
& 0.69 {\rm ln}(\gamma_{\rm max}/\gamma_{\rm min}) &(\alpha=0.5)\nonumber\\
& 13 \gamma_{\rm min}^{-0.5} ~~~~~~~~~~~~&(\alpha=0.75)\nonumber\\
& 74 \gamma_{\rm min}^{-1} ~~~~~~~~~~~~~&(\alpha=1.0).
\label{eq:gu}
\end{eqnarray}
We can invert (\ref{eq:urat}) to determine an ``equipartition brightness temperature,''
\begin{equation}
T_{\rm eq} \approx 10^{11} [D_{\rm Gpc} \nu_{\rm m,GHz} \Theta_{\rm mas}(1+z)^{-5} \delta^3
g^{-1}_{\rm u}(\alpha,\gamma_{\rm min},\gamma_{\rm max})]^{1/(7+2\alpha)}~~{\rm K}.
\label{eq:Teq}
\end{equation}
Features with brightness temperatures higher than this have energy densities dominated
by relativistic particles.

Use of this method requires very accurate measurements of the
angular size $\Theta$, spectral turnover frequency $\nu_{\rm m}$, and flux density
at the turnover $F_{\rm m}$ of individual
features in jets, a difficult undertaking requiring multifrequency VLBI observations.
With current VLBI, restricted to Earth-diameter baselines at high radio frequencies,
only a relatively small number of well-observed knots have
properties amenable to reliable estimates of the physical parameters. 
Application of this method to semi-compact jets (between parsec
and kiloparsec scales, observed at frequencies less than 1 GHz), indicates that
jets are close to equipartition on these scales \cite{R94}. The same may
be true for the parsec-scale cores, although some cores and individual very compact knots
can have particle energy densities greatly exceeding $B^2/(8\pi)$ \cite{HKL06,M07}.

\subsection{Inverse Compton X-ray and $\gamma$-ray Emission}
\label{sec:IC}
Derivation of the magnetic field and $N_0$ allows the calculation of the flux of
high-energy photons from inverse Compton scattering\index{inverse Compton scattering}. If the ``seed'' photons that
the relativistic electrons scatter are produced by synchrotron radiation
from the same region of the jet, we refer to the process as ``synchrotron self-Compton\index{synchrotron
self-Compton}" (SSC) emission. In this case, we can apply the same analysis as in
Sect. \ref{sec:E} to obtain the flux density at photon energy $h\nu$ in keV, where
$h$ is Planck's constant:
\begin{eqnarray}
F_\nu^{SSC} &\approx& c_{\rm ssc}(h\nu)_{\rm keV}^{-\alpha}
\Theta_{\rm mas}^{-2(3+2\alpha)}
\nu_{\rm m,GHz}^{-(5+3\alpha)} F_{\rm m,Jy}^{2(2+\alpha)}
\big({{1+z}\over{\delta}}\big)^{2(2+\alpha)}{\rm ln}
\big({{\nu_{\rm max}}\over{\nu_{\rm m}}}\big)~~\mu{\rm Jy}\nonumber\\
&\approx& (c_{\rm tc}T_{\rm m,11})^{3+2\alpha} (h\nu)_{\rm keV}^{-\alpha} F_{\rm m,Jy} 
\nu_{\rm m,GHz}^{1+\alpha} \big({{1+z}\over{\delta}}\big)^{2(2+\alpha)}~\times\nonumber\\
&&~~~~~{\rm ln}\big({{\nu_{\rm max}}\over{\nu_{\rm m}}}\big)
~~\mu{\rm Jy}.
\label{eq:FIC}
\end{eqnarray}
The value of $c_{\rm ssc}$ is 130, 43, 18, and 91 and that of $c_{\rm tc}$ is 0.22,
0.14, 0.11, and 0.083 for $\alpha =$ 0.25, 0.5, 0.75, and 1.0, respectively.
The formula applies over the frequency range $\sim \gamma_{\rm min}^2\nu_{\rm m}$
to $\sim \gamma_{\rm max}^2\nu_{\rm max}$ over which the SSC spectrum is
a power law with slope $-\alpha$. Here, $\nu_{\rm max}$ is the highest frequency at
which the synchrotron spectrum possesses a power-law slope of $-\alpha$.

A simpler formula relates the SSC flux density at frequency $\nu^{\rm c}$ to the
synchrotron flux density at frequency $\nu^{\rm s}$:
\begin{equation}
{{F_\nu^{\rm c}(\nu^{\rm c})}\over{F_\nu^{\rm s}(\nu^{\rm s})}} \approx
c_{\rm cs}(\alpha) RN_0 {\rm ln}{{\nu_2}\over{\nu_{\rm m}}}
\big({{\nu^{\rm s}}\over{\nu^{\rm c}}}\big)^\alpha ,
\label{eq:FcFs}
\end{equation}
where $c_{\rm cs}(\alpha) = 2.7\times 10^{-22},~3.2\times 10^{-19},~3.9\times 10^{-16},
~5.0\times 10^{-13}$ for $\alpha =$ 0.25, 0.5, 0.75, and 1.0, respectively.

The source of seed photons that are scattered may also lie outside the jet or in
some other section of the jet. If the seed photons are monochromatic with
frequency $\nu_{\rm seed}$ and energy density $u_{\rm seed}$, both measured
in the rest frame of the host galaxy, the ``external Compton'' flux density is given by
\begin{equation}
F^{\rm ec}_\nu = c_{\rm ec}(\alpha) {{u_{\rm seed}}\over{\nu_{\rm seed}}} D_l^{-2} N_0 fV
\delta^{4+2\alpha} (1+z)^{1-\alpha} \big({{\nu_{\rm seed}}\over{\nu_{\rm ec}}}\big)^{\alpha},
\label{eq:EC1}
\end{equation}
where $c_{\rm ec}(\alpha) \equiv (c\sigma_{\rm t}/2)(mc^2)^{-2\alpha}$, with
$\sigma_{\rm t}$ being the Thomson cross-section, is tabulated in \cite{JM04}.
If the seed photon distribution is actually a Planck function, as is the case
for the cosmic microwave background [CMB, $u_{\rm seed} = 4.2\times 10^{-13}(1+z)^4$],
we can use the same expression if we multiply it by $c_{\rm planck}(\alpha) =$
0.46, 0.58, 0.75, and 1.00 for $\alpha =$ 0.25, 0.5, 0.75, and 1.0, respectively.
Since the CMB seed photons are only of relative importance in extended jets and
lobes, we express the EC/CMB flux density in units appropriate to the larger scales:
\begin{eqnarray}
F^{\rm ec}_\nu({\rm CMB}) = &{{c_{\rm cmb}(\alpha)Af}\over{1+k}}
g(\alpha,\gamma_{\rm min},\gamma_{\rm max})^{-1} D_{\rm Gpc} \Theta_{\rm as}^3
B_{\mu{\rm G}}^2 \delta^{4+2\alpha}~\times\nonumber\\
&(1+z)^{-2}(h\nu_{\rm ec})_{\rm keV}^{-\alpha}~~{\rm nJy},
\label{eq:ECCMB}
\end{eqnarray}
where  $c_{\rm cmb}(\alpha) \equiv 0.063(4.2\times 10^6)^{-\alpha}c_{\rm planck}(\alpha) =$
$6.4\times 10^{-4}$, $1.8\times 10^{-5}$, $5.1\times 10^{-7}$, $1.5\times 10^{-8}$
for $\alpha =$ 0.25, 0.5, 0.75, and 1.0, respectively.

\subsection{Matter Content}
\label{sec:mat}

A major, yet unsettled question is whether jet plasmas are composed of
``normal matter''---electrons and protons---electrons and positrons, or a combination
of the two. Theoretically, a jet (or sheath of a jet) produced by a wind blown off
the accretion disk should contain mainly normal matter. However, in jets that
are Poynting flux\index{Poynting flux} dominated might be populated by electron-positron pairs
created by interaction of high-energy photons with the electromagnetic fields
or particles in the jet.

While direct detection of the 511 keV electron-positron two-photon annihilation
line would be the strongest evidence for a pair plasma, this emission line is
difficult to detect. One of the problems is that current detectors at this
energy are relatively insensitive. Another is the relatively low cross-section
for annihilation, so that a strong line requires a density higher than that found
in jets. In addition, the line width from such a high-velocity environment as a
jet flow can cause the line to be broadened beyond recognition. If the jet material
mixes with interstellar clouds, however, the positrons can become thermalized and
annihilate with cold electrons, leading to a narrow line. This might be occurring
in the radio galaxy 3C~120, but even in this case observations have failed thus
far to detect a significant spectral feature at the expected energy \cite{M07}.

Interpretation of circular polarization in the core of the quasar 3C~279 suggests
that the jet in this extreme blazar ($\Gamma > 20$; \cite{J05}) contains a
5:1 ratio of positrons to protons \cite{W98}. This relies on explaining the
circular polarization as conversion of linear to circular polarization in an
optically thick region. Tuning of the amount of Faraday rotation versus
synchrotron self-absorption sets the ratio of positive particle type.
However, there has been a suggestion \cite{RB02} that a suitably turbulent
magnetic field in a 100\% normal-matter plasma could also reproduce the data.

Various other attempts to determine the matter content are less direct, and have
led to contradictory conclusions \cite{CF93,R96,SM00,C03}.

\section{AGN Jets in Context}
\label{sec:context}

Although only a minority of AGN possess bright, relativistic jets, jets play an
important role wherever they are present. A high fraction of the
mass-energy accreted onto the black hole can be transformed into highly collimated,
high-speed outflows in the form of both jets and winds. This probably stems
from the difficulty in squeezing magnetic fields and matter with angular momentum
into small spaces, so plasma squirts out perpendicular to the plane of accretion.
We do not yet understand the physical conditions that decide
whether or not relativistic jets form. The main hint that nature provides is that it
seems to be quite difficult in spiral galaxies, but no generally accepted solution has
yet grown out of this clue.
\begin{figure}
\centering
\includegraphics[height=10cm]{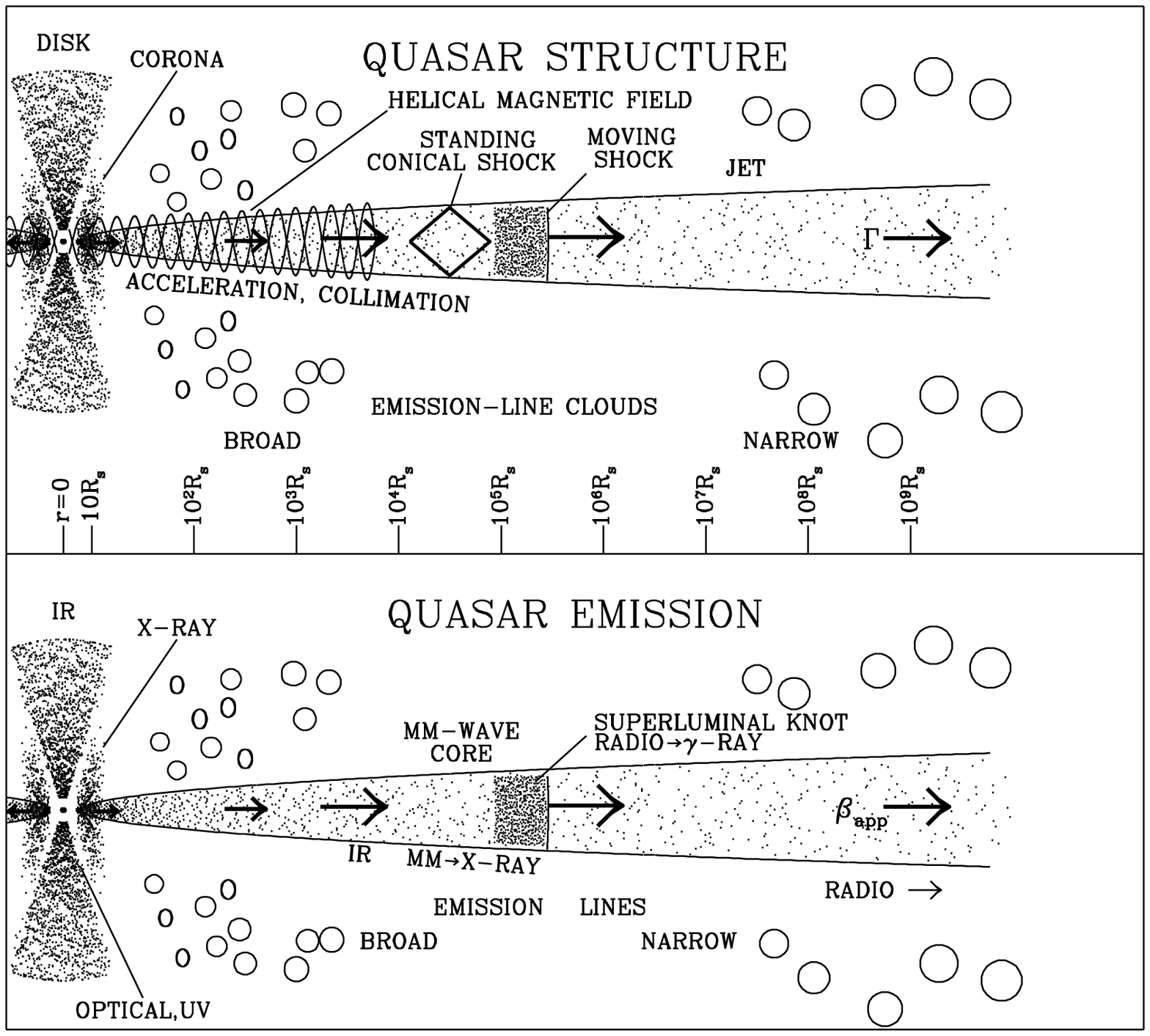}
\caption{Cartoon illustrating the various physical and emission components of a quasar
or other AGN with a relativistic jet. The length scale (shown along the bottom of the top
panel in terms of Schwarzschild radii of the black hole) is logarithmic in order to
include phenomena over a wide range of distances from the black hole. The radiation
produced in the jet is relativistically beamed, while that outside the jet is not}
\label{fig:9}
\end{figure}

Figure \ref{fig:9} sketches the entire AGN system as we currently understand it. The
broad-line clouds are immersed in a more dilute, hotter medium, perhaps in a wind
emanating from the accretion disk. The jet can be thought of as either the same
general outflow, with the velocity increasing toward smaller angles to the polar axes
of the disk. Alternatively, the jet is a distinct flow structure that originates in
the ergosphere of the black hole, surrounded by the wind but with properties quite
different. For example, the jet could be populated mainly by electron-positron pairs,
with normal electron-proton plasma representing a minority of the particles.

Our observational tools for probing jets continue to become more sophisticated with time.
Millimeter-wave antennas in high-Earth orbit as part of a VLBI array can provide
the angular resolution needed to explore regions upstream of the core. Well-sampled
multiwaveband flux and, where possible, polarization monitoring observations from
radio to $\gamma$-ray frequencies might reveal the structure of the jet between
these locations and the black hole. And the ever-surprising blazar jets are likely to
provide us with singular events that unveil yet another clue in our quest to
understand this exotic and enigmatic phenomenon.


\end{document}